\documentclass[twocolumn,RNAAS]{aastex631}

\usepackage{amsmath}

\newcommand{\Jdir}{\ensuremath{\hat{\mathbf{J}}}}
\newcommand{\ea}{\ensuremath{\hat{\mathbf{e}}_1}}
\newcommand{\eb}{\ensuremath{\hat{\mathbf{e}}_2}}
\newcommand{\ec}{\ensuremath{\hat{\mathbf{e}}_3}}
\newcommand{\ei}{\ensuremath{\hat{\mathbf{e}}_i}}

\newcommand{\nexus}{\textsc{nexus+}}

\begin{document}

\title{Early evolution of spin direction in dark matter halos and the effect of the surrounding large-scale tidal field\footnote{\href{}{plopez@unc.edu.ar}Thesis work conducted at Facultad de Matemática, Astronomía, Física y Computación, Universidad Nacional de Córdoba, Córdoba, Argentina.}\footnote{PhD Thesis directed by Manuel Merchán; PhD Degree awarded 2023 July 31.}}

\author[0000-0002-9596-9812]{Pablo López}
\affiliation{Observatorio Astronómico de Córdoba, Universidad Nacional de Córdoba (UNC), Francisco N. Laprida 854, Córdoba, Argentina}
\affiliation{Instituto de Astronomía Teórica y Experimental, CONICET-UNC, Laprida 922, Córdoba, Argentina; \rm \href{mailto:plopez@unc.edu.ar}{plopez@unc.edu.ar}}



\begin{abstract}
It is usually assumed that the angular momentum (AM) of dark matter halos arises during the linear stages of structure formation, as a consequence of the coupling between the proto-haloes' shape and the tidal field produced by their surrounding density perturbations. This approach, known as linear tidal torque theory (TTT), has been shown to make fairly good predictions about the mean evolution of both the AM amplitude and orientation up to approximately the time when the proto-haloes collapse. After this point, proto-haloes are increasingly affected by non-linear processes that are not taken into account by the model. However, it has been seen in numerical simulations that, even at very early stages, the AM of proto-haloes is systematically reoriented towards perpendicularity with respect to the forming cosmic filaments, in contradiction with the fixed direction expected from the TTT. In this work we present a novel analytical approach that introduces an anisotropic scaling factor to the standard TTT equations, which allows the AM orientation to change in time, even during the linear regime. The amplitude and direction of this shift depend on the large scale tidal field around the forming proto-haloes. Our results significantly improve the predictions for the AM direction up to the time of protohalo collapse and, in some cases, even further in time.
\end{abstract}



\section{Introduction} \label{sec:intro}

The present work focuses on the angular momentum evolution of dark matter (DM) halos within the context of the large-scale structure of the Universe. According to the standard cosmological model, galaxies and groups of galaxies form in the potential wells of these structures. DM halos arise from small fluctuations in the density field of the primitive Universe. As they are amplified by gravitational instability and virialize, they also migrate and cluster to form a complex network of filaments, nodes, walls and voids that in the literature is usually known as the cosmic web \citep[][see \citealt{libeskind2018} for a recent revision of this topic]{bondetal1996,weygaert&bond2008,aragoncalvoetal2010,cautunetal2014}. Hence, the study of the spatial distribution, evolution and properties of halos within the large-scale structure of the Universe allows us to connect phenomena on a wide range of scales: from the determination of cosmological parameters to galactic astrophysics.

One of the key ingredients in the models of galaxy formation is their spin, which is usually assumed to be inherited from the angular momentum (AM) of their hosting DM halos. In the current paradigm, the AM arises during the early stages of structure formation, due to the coupling between the shape of the forming proto-halo and the tidal field produced by its surrounding matter distribution \citep{hoyle1951,peebles1969,doroshkevich1970,white1984}. This approach has been the basis of an heterogeneous family of models and predictions that can be summarized as the tidal torque theory (TTT). The TTT predicts the growth of AM in the frame of an isotropically expanding Friedmann-Lema\^{i}tre-Robertson-Walker universe \citep[FLRW,][see \citealt{ellisyvanelst1999} for a complete review]{friedmann1924,lemaitre1931,robertson1935,walker1937}. It adopts a Lagrangian approach, considering the mass elements that conform each proto-halo as individual particles. It also assumes that, before shell crossing occurs, the evolution of the particles can be reasonably well described by the Zel'dovich approximation \citep{zeldovich1970}. 

In this context, if we consider a proto-halo whose moment of inertia with respect to its center of mass is $\mathbf{I}$, located within a tidal field $\mathbf{T}$, the $i$-th component of its AM can be expressed as a function of the cosmic time $t$ through \citep{white1984}:
\begin{equation}
    \label{eq:TTT}
    J_i(t)=a^2(t)\dot{D}(t) \epsilon_{ijk}T_{jl}I_{lk},
\end{equation}
where $a(t)$ is the usual scale factor of the FLRW universe, $D(t)$ is the linear growth factor that describes how density perturbations evolve in time, the dot denotes a derivative with respect to cosmic time and $\epsilon_{ijk}$ represents the fully antisymmetric rank-three tensor (for a more detailed description of the TTT, see e.g. Section 2 in \citealt{lopezetal2019} and references therein). 

Eq. \eqref{eq:TTT} states that, during the early stages of structure formation, the AM is a result of the misalignment between the shape of proto-halos and their surrounding tidal field. This expression provides a framework for calculating statistical properties associated to the AM growth \citep[e.g.][]{hoffman1986,heavensypeacock1988,steinmetzybartelmann1995,sugermanetal2000,leeypen2000,porcianietal2002a,porcianietal2002b} and to compare these predictions with results from numerical simulations. 


One important thing to notice in Eq. \eqref{eq:TTT} is that the time dependence of $J_i(t)$ is determined solely by the isotropic factor $a^2(t)\dot{D}(t)$, which implies that each component of the AM grows at the same rate. Hence, according to this formulation, only the amplitude of the AM is expected to change, whereas its direction should \textit{remain constant over time}, at least during the period in which the model assumptions are valid. This is a key implication of the standard TTT that is often overlooked: the model allows to predict preferred directions for the AM, but it can not address any variation in the spin orientation. In fact, different implementations of this framework that accurately estimate the AM direction with respect to the large-scale structure \citep[e.g.][]{leeypen2000,porcianietal2002b}, and even explain secondary trends such as the mass dependency of the spin-filament alignment \citep{codisetal2015}, usually ignore the time factor and only focus on the form of the tensors $\mathbf{I}$ and $\mathbf{T}$ or the correlation between them. 

\section{Evolution of the spin-filament alignment}
\label{sec:spin-LSS}

\begin{figure}[ht!]
\includegraphics[width=\columnwidth]{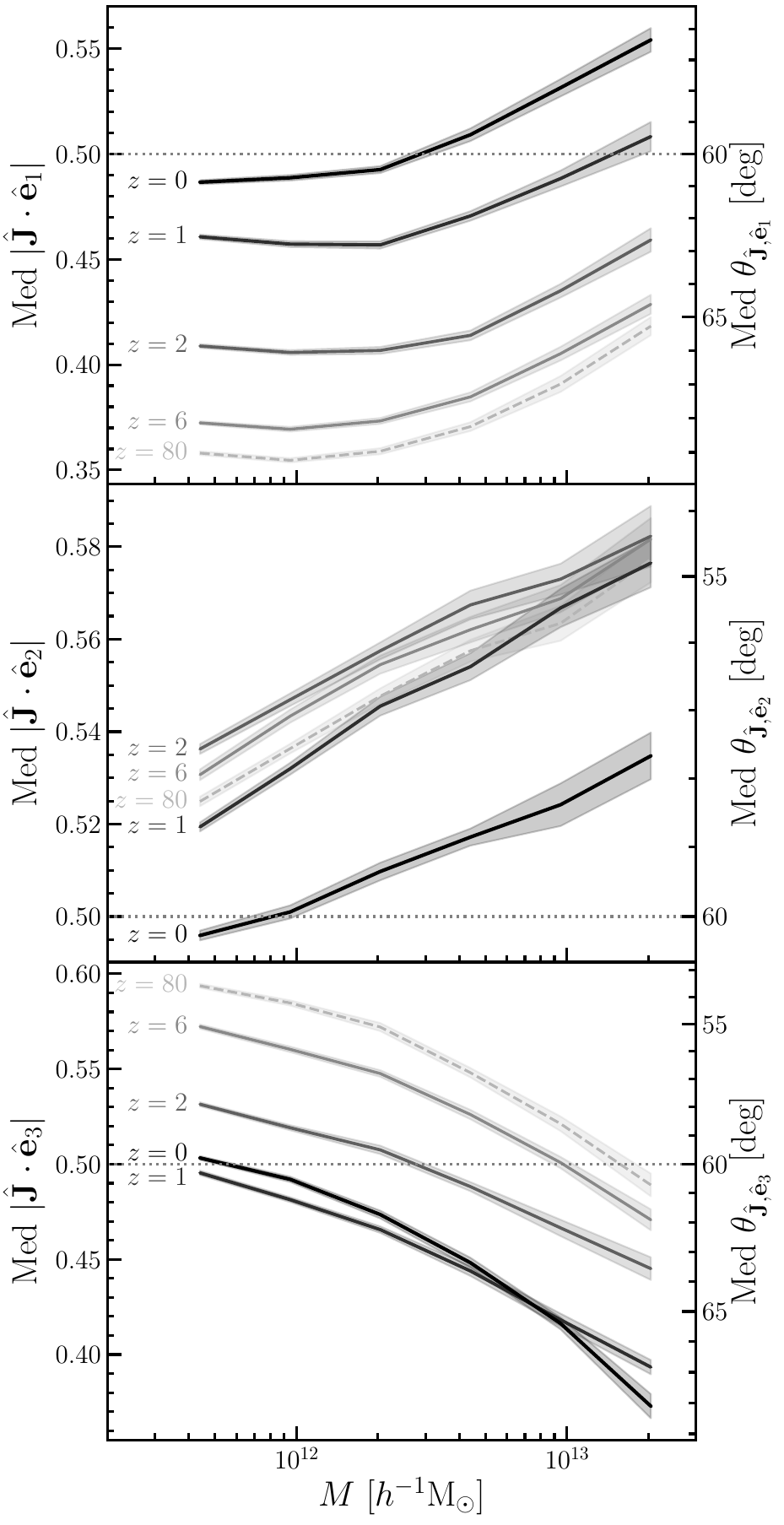}
\caption{Median alignment between the AM of halos at different redshifts and the three preferred directions associated to their embedding filaments at present time. The indexes $i=1,2$ represent, respectively, the first and second axes of collapse, whereas $i=3$ indicates the direction of expansion, i.e. the spine of the filament. Each curve corresponds to a different redshift, as indicated by the label. The median alignment is ploted as a function of halo mass. Figure taken from \citealt{lopezetal2021}.
\label{fig:evol_Je_paper2}}
\end{figure}

In the first part of this thesis we use DM only cosmologial simulations to show that, despite the prevailing assumption, the spin direction of proto-halos undergoes significant evolution even during the early stages, systematically shifting towards perpendicularity with respect to the developing cosmic filaments. This contradicts the fixed direction expected from Eq. \ref{eq:TTT} and is the main result of the first part of the thesis.

In order to perform this analysis we determine, on the one hand, the evolution of the AM of the proto-halos between $z=80$ and $z=0$ and, on the other, the main directions of the cosmic filaments in which each halo inhabits at present time. In the first case, after identifying virialized groups at $z=0$ using a Friends of Friends (FOF) algorithm, we follow the particles of each group back in time determining, at each instant $t$, the spin direcion $\Jdir(t)$ of the corresponding Lagrangian patch. In the second case we use the \nexus method \citep{cautunetal2013} to identify cosmic filaments at $z=0$ and determine their principal axes $\ei$, where $i=1,2$ represent the first and second direction of collapse, respectively, and $i=3$ corresponds to the direction of expansion or spine of the filament. Next, we characterize the evolution of the alignment between $\Jdir(t)$ and $\ei$ by computing the median value of the cosine of the angle between them.

In Figure \ref{fig:evol_Je_paper2} we show one of the main results of this procedure. As it can be seen, at $z=80$ the AM has a mass-dependent preferred orientation: whereas for all masses there is a strong trend for $\Jdir$ to be perpendicular to $\ea$, low mass halos are more tipically aligned with $\ec$, while high mass halos are mostly aligned with $\eb$. This is cualitatively consistent with predictions from TTT \citep{leeypen2000,porcianietal2002b}. 

For $z=0$, the trends we observe are different from the early stages, but widely known: the higher the mass of halos, the stronger their AM is perpendicularly aligned with respect to the spine of their embedding filaments. 
While this is not predicted by TTT, the usual interpretation is that certain late non-linear processes, such as mergers, anisotropic secondary accretion, and fly-by's, which strongly correlate with the preferred directions of the large-scale structure, affect the AM evolution that was produced by tidal torques during the early stages and are ultimately responsible for the present day alignments. However, it is clear that, at least with respect to $\ea$ and $\ec$, there is a significant change in the spin direction from $z=80$ to $z\sim1$, i.e. an important fraction of the total shift in orientation is produced when the the assumptions of TTT are valid. Moreover, this shift is systematic and almost mass-independent: the AM becomes progressively perpendicular to $\ec$ and tend to align with $\ea$.

This results suggest that the TTT might be extended or modified in order to account for the early change in spin orientation, and that the direction of the large-scale structure must play a key role in understanding how and why this variation occurs.

\section{Modified TTT}
\label{sec:TTT_mod}

\begin{figure*}[ht!]
\includegraphics[width=2\columnwidth]{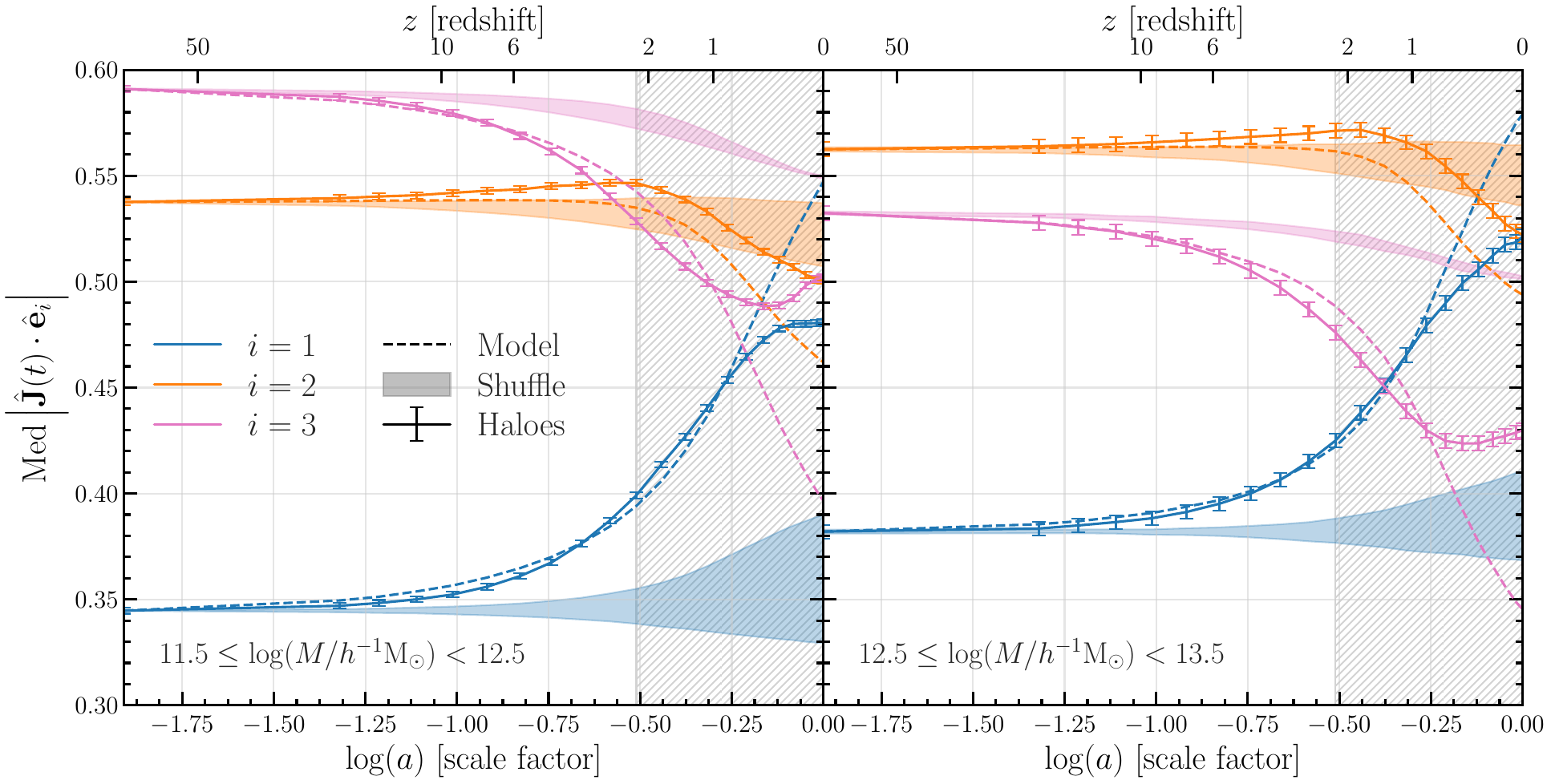}
\caption{Evolution of the alignment of the AM with respect to the three preferred directions, $\ei$, with $i=1,2,3$, associated to their embedding filaments at $z=0$. The solid lines show the median value of the AM for DM halos in our simulation, with the errorbars representing the $3\sigma$ bootstrap error. The dashed lines correspond to the median value predicted by the model expressed by Eq. \ref{eq:TTT_mod}. The coloured shaded areas represent the $3\sigma$ dispersion of the predictions of the model for $10$ realizations with halos and large-scale tidal fields randomly shuffled.
\label{fig:evol_Je}}
\end{figure*}

Motivated by the above results, the next part of the thesis focuses in a reformulation of the TTT that allows early time variation of the spin direction. First, notice that a finite-volume region, such as the one in which a given proto-halo and its surrounding tidal field evolve, can suffer from the effects of perturbations on even larger scales \citep{akitsuetal2017}. These fluctuations, despite having small amplitudes, are capable of affecting the evolution of structures at smaller scales due to the non-linear coupling of modes corresponding to different wavelengths. 

As noted by \citet{schmidtetal2018}, there are two leading effects that come to play in this kind of process. On the firsts place, when the large-scale perturbation is associated with the presence of a coherent over- or underdensity, which results in a higher or lower local expansion rate relative to the global expansion. Secondly, when a significant tidal force can be associated with the long-wavelength fluctuation, which produces an increase in the anisotropy of the local statistics. The last scenario has been analyzed in detail by \citet{schmidtetal2018}, who showed that the evolution of a small region embedded in a large-scale tidal field can be described in terms of a modified set of cosmological parameters, in a similar way as in the ``separate universe'' picture \citep{salopekybond1990,wandsetal2000}. 

More concretely, due to the presence of a large-scale tidal field, the region of interest will evolve differently from the larger, isotropically expanding background. Whereas the universe grows according to the standard scale factor $a(t)$, our region will expand anisotropically. Moreover, within a coordinate system rotated into the principal axes of the large-scale tidal field, this evolution can be described simply by three scale factors, $a_i$, with $i=1,2,3$. 

At this point is useful to consider the expansion factor ratios $\alpha_i=a_i/a$. These can be defined such that $\alpha_i\rightarrow1$ at the limit of early times ($a\rightarrow0$). As shown by \citet{stuckeretal2018} and \citet{schmidtetal2018}, under certain reasonable assumptions, the evolution of $\alpha_i$ can be approximated during the linear and quasi-linear regime by:
\begin{equation*}
\alpha_{i}(t) = 1-D(t)\lambda_i,
\end{equation*}
where $\lambda_i$ are the eigenvalues of the large-scale tidal field. 

Hence, in a frame aligned with the surrounding large-scale tidal field, the evolution of the AM of a proto-halo can be expressed as:
\begin{equation}
\label{eq:TTT_mod}
J_i(t) \propto a^2(t)\dot{D}(t)\alpha_{j}(t)\alpha_k(t)\epsilon_{ijk},
\end{equation}
where is the value of the $i$-th component of the AM at a very early epoch, e.g. at the initial conditions of the simulation. This equation is the most important result in the thesis.

As an example, let us consider a proto-halo that evolves in a region that will end up forming a cosmic filament. In this case, the large-scale tidal field is typically characterized by two directions of collapse and a third direction of expansion. The AM growth in the latter direction corresponds to $i=3$ and, according to Eq. \eqref{eq:TTT_mod}, it is given by:
\begin{equation*}
J_3(t) \propto a^2(t)\dot{D}(t)\alpha_1(t)\alpha_2(t).
\end{equation*}
Since $i=1,2$ represent the two directions of collapse, we expect $\alpha_{1,2} < 1$ and, therefore, that $J_3(t)$ will evolve at a lower rate than it is expected from the standard implementation of the TTT. In opposition, the corresponding expressions for $J_1(t)$ and $J_2(t)$ depend on $\alpha_3 > 1$, so it is likely that they will grow at a higher rate. Thus, a rough first look at this model tells us that \textit{the AM direction of a proto-halo that grows within a filament-to-be region will shift towards perpendicularity with respect to the spine of the filament}.

Finally, in order to test this model, we implement Eq. \eqref{eq:TTT_mod} on a simulated population of DM halos and compare the result with the true evolution of their AM. For this, we first calculate the large-scale tidal tensor around each proto-halo at $z=80$ and determine the corresponding eigenvalues and eigenvectors. Since we do not know, a priori, which long-wavelength mode effectively couples with the region of interest, before computing the large-scale tidal tensor we smooth the density field around each protohalo using gaussian kernels of different radii. Hence, we choose the smoothing scale for which the predictions minimize a given error function with respect to the true value of the AM. To check that this method does not bias our results, we repeat the procedure $10$ times by randomly shuffling the proto-halos and their tidal fields, thus producing a sample of predictions where the goodness of fit can only be spurious.

In Figure \ref{fig:evol_Je} we present the results of this procedure. The left (right) hand panel shows the median evolution of the alignment between $\Jdir$ and $\ei$ for low (high) mass halos. Solid lines represent the true AM measured in the simulation and the error bars indicate the $3\sigma$ bootstrap error, while the dashed lines represent the evolution predicted by Eq. \eqref{eq:TTT_mod}. The coloured dashed regions indicate the $3\sigma$ dispersion of the values obtained with the randomly shuffled samples. 

Now, both at low and high masses, there seems to be a good agreement between the measured evolution and the prediction of our model during the linear and quasi-linear regime, specially with respect to $\ea$ and $\ec$. Recalling that the standard implementation of the TTT produces a constant direction, that is, horizontal lines, Eq. \eqref{eq:TTT_mod} yields better results up to about $z\sim2$, although in some cases the agreement seems to extend even further in time. Notice that the shuffled samples are not able to reproduce this, hence probing that the method is not biasing our results and suggesting that, indeed, the accuracy of the prediction is due to the physical connection between the AM and the large-scale tidal field encoded in Eq. \eqref{eq:TTT_mod}. 

\section{Summary}
\label{sec:summary}

As a general summary, in this thesis we first use N-body cosmological numerical simulation to show that there is a systematic variation of the AM direction of DM halos, even during the early stages of structure formation (Fig. \ref{fig:evol_Je_paper2}) and that this variation can not be understood in terms of the standard implementation of the TTT (Eq. \ref{eq:TTT}). Next, we present a novel analytical approach that adopts the ``separate universe'' method in order to introduce an anisotropic scaling factor to the standard TTT equations (Eq. \ref{eq:TTT_mod}). This allows the AM orientation to change in time, even during the linear regime. The amplitude and direction of this shift depend on the large-scale tidal field around the forming proto-halos. Finally, we test our model using numerical simulations. Our results significantly improve the predictions for the AM direction up $z\sim2$ and, in some cases, even further in time. 

\bibliography{PASPsample631}{}
\bibliographystyle{aasjournal}



\end{document}